\documentclass[floats,prb,aps,twocolumn]{revtex4}
\usepackage{graphicx}
\begin{document}

\title{Superconducting  photonic crystals}

\author{Oleg L. Berman$^{1}$, Yurii E. Lozovik$^{2}$, Sergey L. Eiderman$^{2}$ and Rob D. Coalson$^{3}$}

\affiliation{\mbox{$^{1}$Department of Physics and Astronomy,
University of Pittsburgh,}  \\ Pittsburgh, PA 15260 USA
 \\ \mbox{$^{2}$ Institute of Spectroscopy, Russian Academy of
Sciences,}  \\ 142190 Troitsk, Moscow Region, Russia
\\ \mbox{$^{3}$
Department of Chemistry, University of Pittsburgh,}   \\
Pittsburgh, PA 15260, USA }


\begin{abstract}

The band structure of a novel type of photonic crystal with
superconducting constituent elements is calculated numerically via
a plane wave expansion. The density of states and the dependence
of the width of the photonic gap on the filling factor is analyzed
for a two-dimensional photonic crystal consisting of an infinite
array of parallel superconducting cylinders.

\vspace{0.1cm}



\end{abstract}

\maketitle



Photonic crystals, artificial media with a spatially periodical
dielectric function that were first discussed by
Yablonovitch\cite{Yablonovitch} and John\cite{John}, are the
subjects of growing interest due to various modern
applications.\cite{Eldada,Chigrin} This periodicity can be
achieved by embedding a periodic array of constituent elements
(``particles'') with dielectric constant $\varepsilon_{1}$ in a
background medium characterized by dielectric constant
$\varepsilon_{2}$. Different materials have been used for these
elements: dielectrics, semiconductors and
metals.\cite{Joannopoulos1,Joannopoulos2,Sun_Jung,Sun,Maradudin,Kuzmiak}

Photonic gaps are formed at frequencies $\omega$ for sufficiently
high dielectric contrast $\omega^{2}(\varepsilon_{1}(\omega) -
\varepsilon_{2}(\omega))$, because the quantity
$\omega^{2}\varepsilon(\omega)$ enters in the wave
equation.\cite{Joannopoulos1,Joannopoulos2} Thus, only
metal-containing photonic crystals can maintain the necessary
dielectric contrast at small frequencies due to Drude-like
behavior $\varepsilon_{Met} (\omega) \sim
-1/\omega^{2}$.\cite{Maradudin,Kuzmiak}   But the damping of
electromagnetic waves in metals can suppress many useful
properties of metallic photonic crystals.

We suggest in this Letter a novel type of photonic crystal
consisting of  {\it superconducting} elements (embedded in
dielectric media) or inverse superconducting crystal with (vacuum)
holes. This photonic crystal can be used for (THz) frequencies
smaller than superconducting gap where the damping is negligible.

For simplicity, we consider superconducting particles in the
London approximation (i.e. for London penetration length
$\delta_{L} = [mc^{2}/(4\pi n e^{2})]^{1/2} \gg \xi$, $\xi$ is the
coherence length). The optical properties of superconductors have
been extensively studied (see, for example,
Refs.~[\onlinecite{Abrikosov,Lozovik_prb}] and references cited
therein). At $(T_{c} -T)/T_{c} \ll 1$ and $\hbar \omega \ll \Delta
\ll T_{c}$ a simple relation relation for the current density is
given by\cite{Abrikosov}
\begin{eqnarray}\label{sup_cur}
\mathbf{J}(\mathbf{r}) = \left[-\frac{c}{4\pi\delta_{L}^{2}} +
\frac{i\omega \sigma}{c}\right] \mathbf{A}(\mathbf{r}) ,
\end{eqnarray}
where $\sigma$ is a conductivity of normal metal.

Substituting Eq.~(\ref{sup_cur}) into the wave equation
\begin{eqnarray}\label{subst}
-\nabla^{2}\mathbf{E} = - \frac{1}{c^2} \epsilon_{0}
\frac{\partial^{2}\mathbf{E}}{\partial t^{2}} -
\frac{4\pi}{c^{2}}\frac{\partial \mathbf{J}(\mathbf{r})}{\partial
t}
\end{eqnarray}
and seeking solutions with harmonic time variation of the electric
field, i.e.,  $\mathbf{E}(\mathbf{r},t) =
\mathbf{E}_{0}(\mathbf{r})e^{i\omega t}$, $\mathbf{E} = i\omega/c
\mathbf{A}$, we have for a two-component photonic crystal:
\begin{eqnarray}\label{substparteq}
 && -\nabla^{2}\mathbf{E} =  \frac{\omega^{2}}{c^{2}} \left[\epsilon_{0}
 + \left( \varepsilon_{s} (\omega) - \varepsilon_{0} \right)
 \sum_{\{n_{i}\}}
 \eta (\mathbf{r} \in S)  \right] \mathbf{E} \; ,
\end{eqnarray}
where $\eta(\mathbf{r} \in S) = 1$ if $\mathbf{r}$ is inside
superconductor elements $S$, and otherwise $\eta(\mathbf{r} \in S)
= 0$, $\varepsilon_{s}(\omega)$ being the dielectric function of
the superconductor and $\varepsilon_0$ the dielectric constant of
the surrounding medium. Here we adopt the approximation
$\varepsilon _{s} (\omega) = 1 +
c^{2}/(\delta_{L}^{2}\omega^{2})$, which neglects damping in the
superconductor. The summation in Eq.~(\ref{substparteq}) goes over
all lattice nodes $\mathbf{a}\{n_{i}\} \equiv
\sum_{i=1}^{d}n_{i}\mathbf{a}_{i}$ characterizing positions of
superconducting particles ($\mathbf{a}\{n_{i}\}$ is the coordinate
space vector of a superconducting particle occupying a particular
node of the lattice).

Here we calculate the band structure of a two-dimensional photonic
crystal consisting of an infinite array of superconducting
cylinders that form a square lattice. Specifically, by expanding
the electric field in plane waves (the plane wave expansion method
commonly used for band structure calculations of
solids\cite{Ashcroft} was applied for calculation of the band
structure of 3D photonic crystal with dielectric spheres in the
nodes in Ref.~[\onlinecite{Ho}]) we reduce the Maxwell Equation
(3) for a superconducting photonic crystal to the following
eigenvalue problem:
\begin{eqnarray}\label{maxw}
&& \sum_{\mathbf{G}'} \left( \delta_{\mathbf{GG}'} (\mathbf{k} +
\mathbf{G})^{2} -
\frac{P^2}{c^2}M_{\mathbf{GG}'}\right)E_{k}({\mathbf G'})
\nonumber
\\ && = \frac{\omega^2}{c^2}E_{k}({\mathbf G}) ,
\end{eqnarray}
where $P = c/\delta_{L}$; $\mathbf{G} = i_{1}\mathbf{G}_{1} +
i_{2}\mathbf{G}_{2}$ is the reciprocal lattice vector for 2D
square lattice. Here $i_{1}$ and $i_{2}$ are arbitrary integers,
and $\mathbf{G}_{1}$ and $\mathbf{G}_{2}$ are 2D orthogonal
vectors: if the unit cell vectors $\mathbf{a}_{1}$ and
$\mathbf{a}_{2}$ of the 2D square lattice are oriented along
$\mathbf{\hat x}$ and $\mathbf{\hat y}$ real space unit vectors,
then $\mathbf{G}_{1} = 2\pi \mathbf{\hat x} /a$ and
$\mathbf{G}_{2} = 2\pi \mathbf{\hat y}/a$, with $a$ being the
lattice spacing. Furthermore, $E_{k}({\bf G})$ are the Fourier
components of the electric field eigenfunction $E_k({\bf x})$. The
matrix $M_{\mathbf{GG}'}$ is related to the Fourier transform of
the periodic dielectric function in this superconducting photonic
crystal:
\begin{eqnarray}\label{Four}
\varepsilon ({\bf G} - {\bf G}') &=&  \delta_{\bf{GG}'} +
\frac{P^2}{\omega^2}M_{\bf{GG}'}  ,
\end{eqnarray}
where
\begin{eqnarray}\label{MGG'}
M_{\bf{GG}'} &=& f\varepsilon + (f-1), \ \ \ {\bf G} = {\bf G}'
\nonumber \\
M_{\bf{GG}'} &=& 2 f(\varepsilon -1) \frac{J_{1}(|{\bf G} - {\bf
G}'|r)}{|{\bf G} - {\bf G}'|r}, \ \ \ {\bf G} \ne {\bf G}'.
\end{eqnarray}
Here $f \equiv S_{supercond}/S = \pi r^2/a^2 $ is the filling
factor of the superconductor($S_{supercond}$ is the
cross-sectional area of the cylinder [in the the plane
perpendicular to the cylinder axis]; $S$ is the the total area
occupied by the lattice unit cell, $r$ is the cylinder radius, $a$
is the lattice spacing); finally, $J_{1}$ is a Bessel function.

The band structure for the filling factor $f = 0.3$ and $P = 5
\times 2 \pi c/a$ is presented on Fig.1 (we use as the unit of
frequency the lattice frequency $2\pi c/a$).\cite{foot} For a YBCO
superconductor $\delta_{L} \approx 200 nm$, and thus $P \approx
1.499 \times 10^{15} s^{-1}$. We have picked the value of $P = 5$
in units of $2\pi c /a$ to attain a significant band gap in the
spectrum. Since for YBCO the value of $P$ is fixed, our results
with $P = 5 \times 2 \pi c/a$ correspond to the specific lattice
spacing $a = 6.283 \mu m$. The density of photonic states for the
crystal presented in Fig.2 is defined as the total number of
different wavevectors $k$ in all directions in one unit cell of
the reciprocal lattice corresponding to the dispersion curve
$\omega = \omega(k)$ in a small frequency region divided by the
frequency range spanned. As illustrated in Fig.3, the width of the
photonic gap is a nonmonotonic function of the filling factor $f$.
A photonic gap exists only in the range $0.23 <f < 0.44$. The
widest gap ($\Delta = 0.35$) corresponds to $f=0.32$.

\emph{In summary}, we have obtained the band structure of a
superconducting photonic crystal. The width of the photonic gap
for the system considered is a non-monotonic function of the
filling factor. The advantage of a photonic crystal with
superconducting particles is that the dissipation of the incident
electromagnetic wave due to the imaginary part of the dielectric
function is much greater for normal metallic than for
superconducting particles, because the imaginary part of the
dielectric function for superconducting particles is negligible in
comparison with the imaginary part of the dielectric function for
normal metal particles at frequencies smaller than the
superconducting gap. Thus, in this frequency regime, for a
photonic crystal consisting of several layers of metallic
scatterers the dissipation of the incident electromagnetic wave by
an array of superconducting particles is expected to be less than
that obtained from an analogous array composed of normal metallic
particles. Note that experimental studies of superconducting
metamaterials have been carried out recently.\cite{Ricci}


\section*{Acknowledgements}
Authors are thankful to Hong Koo Kim, David W. Snoke and
participants of 24th International Conference on Low Temperature
Physics (Orlando, Florida, USA, August 2005) for useful and
stimulating discussions.  Yu. E.~L. has been supported by the
INTAS and RFBR grants.  R. D.~C. has been supported by the
National Science Foundation.


\newpage


\newpage

\begin{figure}
\includegraphics[width = 17cm, height = 16cm]{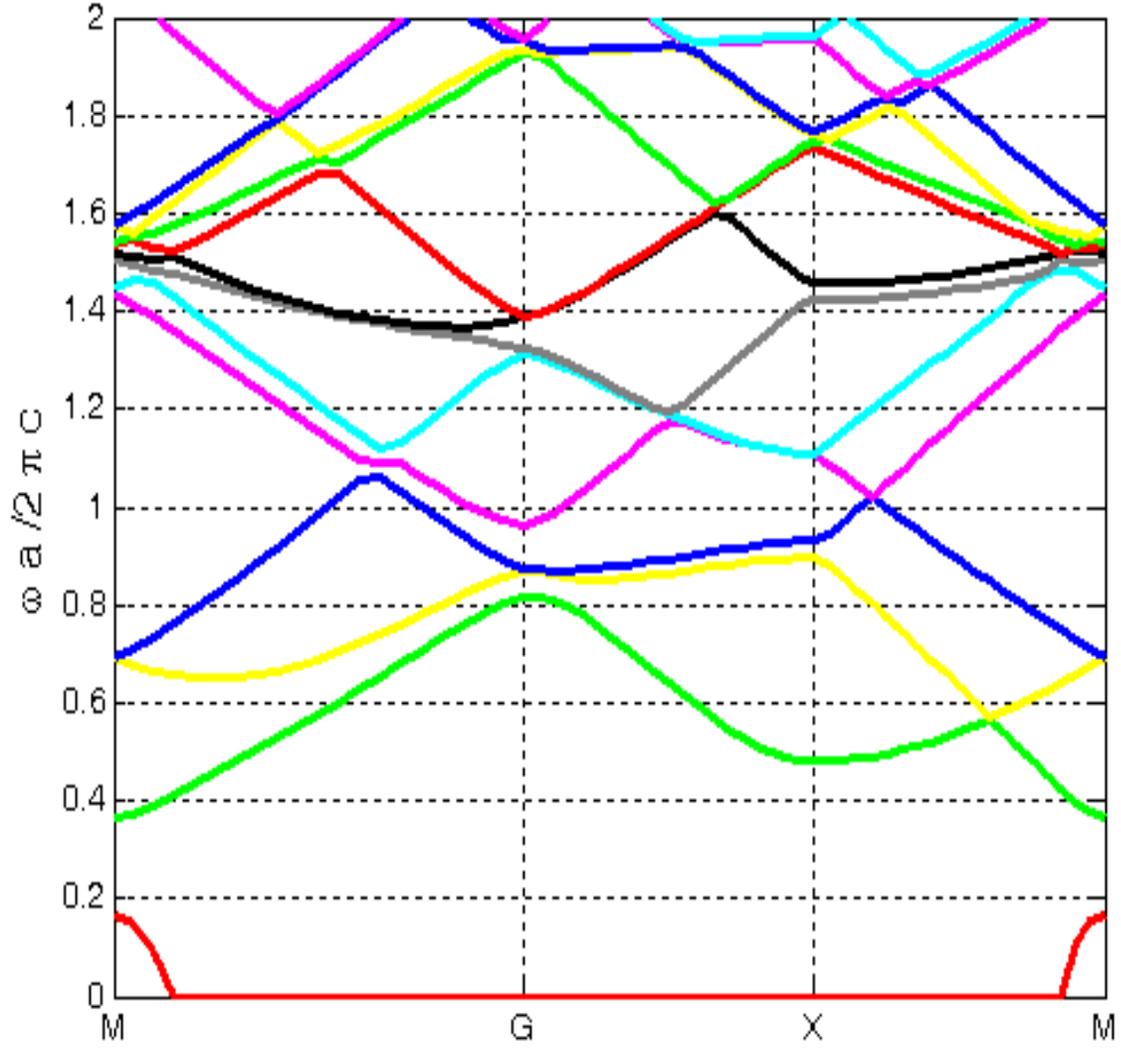}
\caption{Dispersion dependence for a two-dimensional
superconducting photonic crystal with square lattice consisting of
infinite cylinders having circular cross section. E-polarization
is considered, with $f=0.3$ and $P =5  \times 2 \pi c/a$.  The
ordinate plots frequencies in lattice units $2\pi c/a$. A band gap
is clearly apparent. }
\end{figure}

\newpage
\newpage

\begin{figure}
\includegraphics[width = 17cm, height = 16cm]{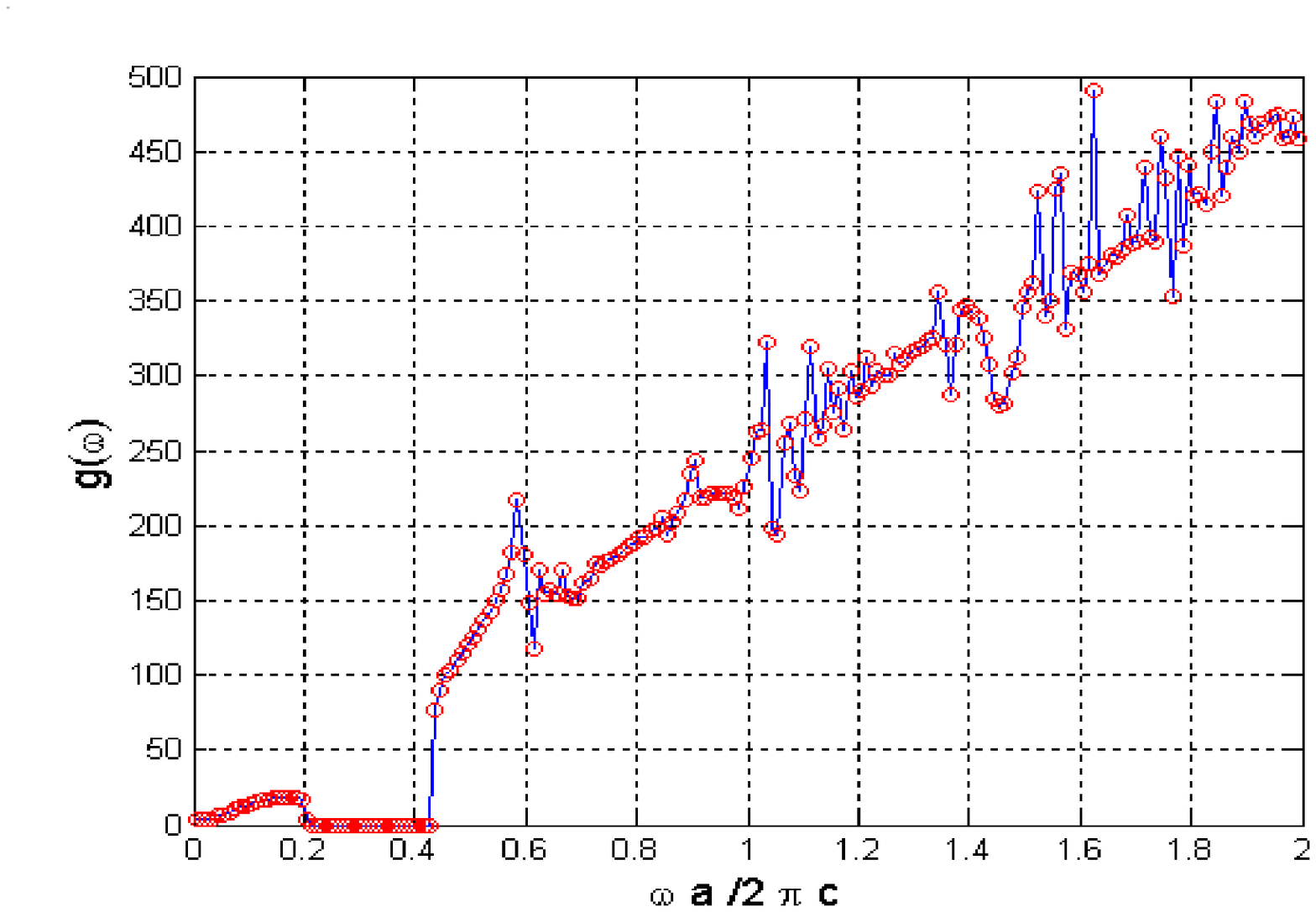}
\caption{The density of states $g(\omega)$ for a two-dimensional
superconducting photonic crystal  with square lattice
(E-polarization) with the filling factor $f=0.3$ and $P=5  \times
2 \pi c/a$. The abscissa plots frequencies in lattice units $2\pi
c/a$.}
\end{figure}

\newpage
\newpage

\begin{figure}
\includegraphics[width = 17cm, height = 16cm]{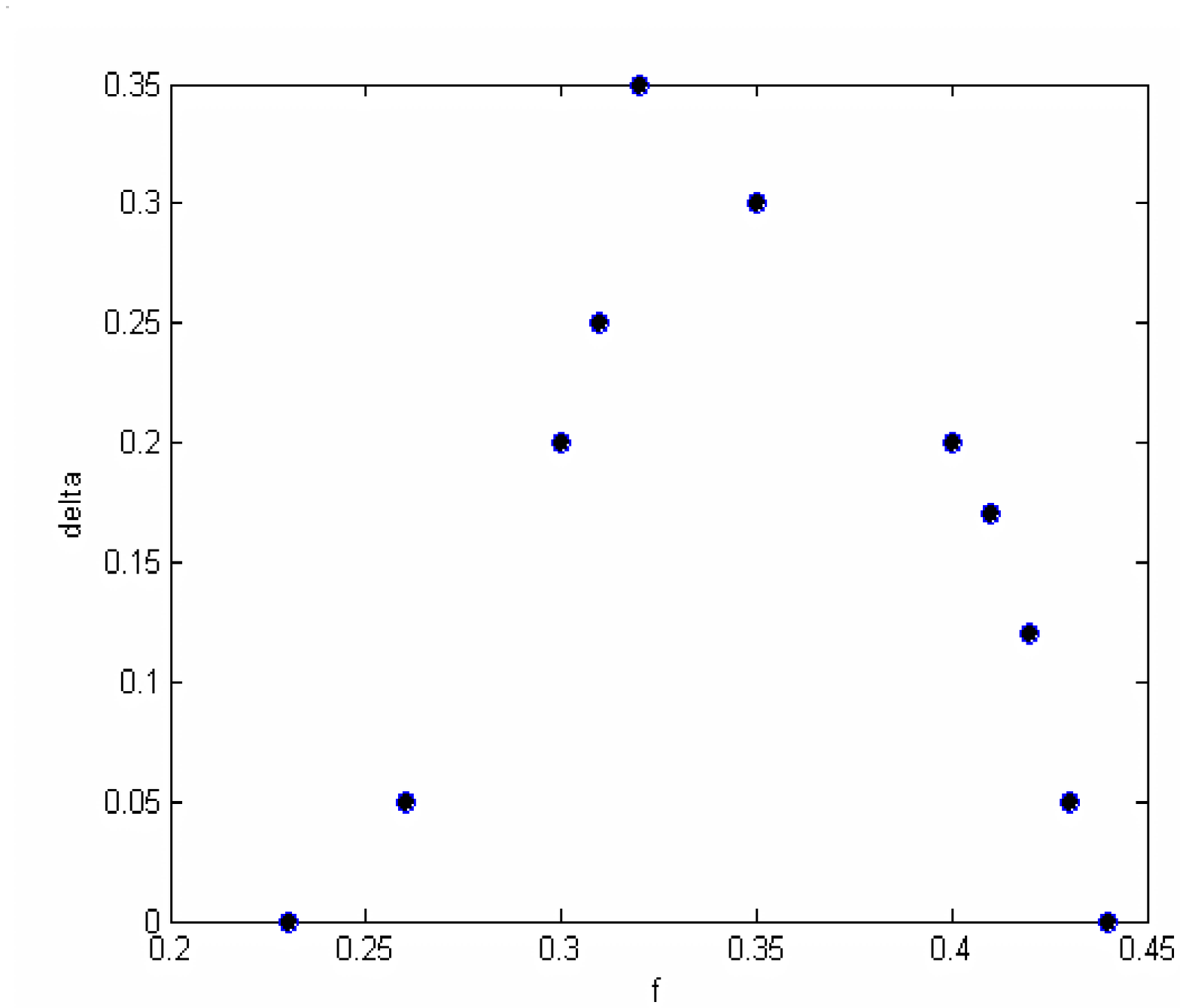}
\caption{Dependence of width of the photonic band on the filling
factor for a two-dimensional superconducting photonic crystal with
square lattice in case of E-polarization ($P = 5  \times 2 \pi
c/a$).  The ordinate plots frequencies in lattice units $2\pi
c/a$.}
\end{figure}


\begin{thebibliography}{999}
\bibitem{foot} We present here only a part of band structure
corresponding to frequencies smaller than superconducting gap. At
much greater frequencies the band structure reduces to that
corresponding to a photonic crystal with normal metal constituent
elements (see Refs.~[\onlinecite{Maradudin,Kuzmiak}]).
\end{thebibliography}
\end{document}